\documentclass[10pt,twocolumn,a4paper]{esaAI}

\title{Spacecraft inertial parameters estimation using time series clustering and reinforcement learning}

\def\authorEmail{K.Platanitis@cranfield.ac.uk}

\author[1]{Konstantinos Platanitis\thanks{Corresponding author. E-Mail: \authorEmail}}
\author[1]{Miguel Arana-Catania}
\author[2]{Leonardo Capicchiano}
\author[1]{Saurabh Upadhyay}
\author[1]{Leonard Felicetti}
\affil[1]{Cranfield University, Cranfield, UK}
\affil[2]{Coactum SA, Bex, CH}

\begin{document}

\makeCustomtitle

\begin{abstract}
This paper presents a machine learning approach to estimate the inertial parameters of a spacecraft in cases when those change during operations, e.g. multiple deployments of payloads, unfolding of appendages and booms, propellant consumption as well as during in-orbit servicing and active debris removal operations.
The machine learning approach uses time series clustering together with an optimised actuation sequence generated by reinforcement learning to facilitate distinguishing among different inertial parameter sets. The performance of the proposed strategy is assessed against the case of a multi-satellite deployment system showing that the algorithm is resilient towards common disturbances in such kinds of operations.

\end{abstract}

\section{Introduction}
The identification of system parameters is a prerequisite for efficient control laws, both in the case of classical state feedback control as well as in modern control techniques\cite{Sidi_1997}. Typical approaches to solve this problem include state observers of an extended state vector that includes the parameters, based on Kalman filtering \cite{ljung1979asymptotic,felicetti2014adaptive,ortega2015parameter}, and other well-studied solutions such as particle filtering \cite{kantas2015particle} or Bayesian methods \cite{TULSYAN2013516}.

While machine learning techniques in conjunction with analytical techniques have been proven to work in generic cases of unknown parameters \cite{nolan2021machine,beintema2021nonlinear}, in general they neither require nor take advantage of prior information that would restrict the search space for the unknown values. This is the case for the deployment of multiple payloads, the unfolding of appendages and booms, propellant consumption, and in-orbit servicing and active debris removal operations.  In this paper, we apply a recently-proposed methodology \cite{sontakke2021causal,mcdonell2023advanced} to the problem of inertial parameter estimation in the particular case of spacecraft that, depending upon mission status, have a distinct set of possible parameters. This includes spacecraft with reconfigurable mechanisms, such as extendable appendages and the particular case of orbital payload deployers whose inertial parameters change with every payload ejection.

In these cases, and under normal operating conditions, the inertial tensor belongs to a predefined set of possible values which depends on the mission status (i.e. appendage deployed/retracted, payload ejected/still onboard). The proposed approach takes advantage of this prior knowledge and, by limiting the search space to a finite set of possible values for the inertia tensor, transforms the generic problem of inertial parameter estimation into a problem that may be solved with data-driven machine-learning techniques.

The motivation behind this work is to present a viable parameter estimation alternative, that works both concurrently with and independently of the commonly employed approaches. Additionally, it presents several advantages compared to traditional approaches: 
\begin{itemize}
    \item The data-driven aspect allows for this approach to be used with any data source, either a priori available telemetry or simulation-generated datasets.
    \item The time-series classification does not require or depend on the knowledge of the exact system dynamics of the disturbance and noise sources, as long as the effects thereof are present in the training dataset.
    \item The RL optimisation technique used allows for future expansion, as for this work the thruster firing optimisation scope has already been augmented from pure end-result to a multi-objective target of minimising the fuel consumption as well. 
\end{itemize}

\section{Methodology}

The proposed method identifies the system's parameters by examining the response under actuation as a time series. To achieve this, we apply the same actuation to a set of systems with different inertial parameters and train a time-series clustering classifier with the dynamics responses. Then, applying the same actuation profile to an unknown (w.r.t. inertial parameters) time series, the classifier is able to distinguish to which of the system models identified during the training phase the response belongs to.

Since not all actuation profiles would produce responses that are adequately different for the classifier to train on, it is important to identify a profile that yields good results. In this work, the actuation profile is considered as a series of discrete pulses of specific length, and the classifier's accuracy is used as a metric to evaluate a particular sequence. Since this is a highly non-convex optimisation problem, we employ reinforcement learning to find optimal actuation sequences to maximise the previous metric.

In this chapter, the representation of a spacecraft is discussed in \ref{dynamics_modelling} along with some nuances regarding actuation modelling. Next, in \ref{tsc_section} the time-series clustering is introduced along with the application in this case. Finally, in \ref{rl_section} the optimisation of the actuation sequence is discussed by use of reinforcement learning. In particular, the Proximal Policy Optimization (PPO) is used to this effect.

\subsection{Dynamics modelling}
\label{dynamics_modelling}
For the dynamics of a spacecraft's attitude, the model used is based on the Euler equations \cite{kane1983spacecraft}:

\begin{equation}
  I\dot{\omega} = -\omega \times (I\omega) + M
\label{eq:attitude_dynamics}
\end{equation}
where $\omega$ is the angular velocity, $I$ is the inertia tensor, and $M$ is the total applied torque. For simulation purposes, a state space model  is used:

\begin{align}
\begin{split}
    \dot{\bar{x}} &= f(\bar{x},\bar{u}) + {\delta}f(\bar{x},\bar{u})\\
    \bar{y} &= h(\bar{x},\bar{u}) + {\delta}h(\bar{x}, \bar{u})
    \label{eq:state_space_representation_eq}
\end{split}
\end{align}
where $\bar{x}=\bar{\omega}$, $f$ is defined as per \cref{eq:attitude_dynamics}, $h(\bar{x},\bar{u})=\bar{x}$, and ${\delta}f$ and ${\delta}h$ represent uncertainties, disturbances, and noise in the dynamics and measurement. Here, the assumption is made that the state of the spacecraft can be inferred from the available sensors, as the state timeseries is necessary to train the classifier.

Moreover, choosing to study the case where the spacecraft only has thrusters available for attitude control, we can model $\bar{u}$ as follows

\begin{equation}
    \bar{u} = 
    \begin{bmatrix}
        M_{1x} & M_{2x} & ... & M_{Nx}\\
        M_{1y} & M_{2y} & ... & M_{Ny}\\
        M_{1z} & M_{2z} & ... & M_{Nz}
    \end{bmatrix}
    \begin{bmatrix}
        u_1 \\ u_2 \\ ... \\ u_N
    \end{bmatrix}
    = A \tilde{u}
\label{eq:actuation_implementation}
\end{equation}
where $M_{ij}$ is the torque impacted along the $j$ axis by the $i$-th thruster, and $u_i \in [0,1]$ is the percentage of the $i$-th thruster's Pulse Width Modulation (PWM)\cite{Sidi_1997} setting.

To reflect system uncertainties during actuation, the assumption is made that gas thrusters have a normally distributed response time due to their mechanical nature. Thus, any commanded PWM duty cycle is intercepted and recalculated with noise added to the signal, to imitate the variance in the response of the actual system. Measurement noise will be based on performance values from typically available sensors such as accelerometers and IMUs \cite{Kuga2013}, and will be added to the actual state values and used as the observable state, which will then be fed to the classifier for training.

\subsection{Time Series Clustering}
\label{tsc_section}
Time series clustering is an unsupervised machine learning technique for parsing multiple time series datasets and grouping them based on their similarity \cite{aghabozorgi2015time}. Here we use the $k$-means algorithm for this purpose. Common metrics used to measure the similarity of time series are the Euclidian metric, the Dynamic Time Warp Barycenter Adjustment (DBA) metric \cite{DTWpaper,DBApaper}, or the Soft-DTW \cite{SoftDTWpaper} metric. The latter two approaches are used in this case, as they allow for better clustering performance in cases where the individual time series belonging to a single cluster are possibly shifted in time. This metric would accommodate for variable signal propagation delays that would not work well with pure Euclidian distance metrics. The time series analysis in the simulations was performed using the tslearn library\footnote{\url{https://github.com/tslearn-team/tslearn}} \cite{tslearn}, which implements the $k$-means algorithm and all of the aforementioned metrics.

The proposed methodology involves applying time series clustering to the dynamics of a spacecraft (i.e. $\omega_x, \omega_y, \omega_z$) when a particular thruster pulse sequence has been applied. Since the dynamics response of the same spacecraft will vary based on the actual inertia tensor, it is thus feasible to train a classifier to learn these different response profiles and subsequently identify which inertia tensor results in a particular response, given the same actuation profile.

The performance and accuracy of a trained classifier can be examined by statistical measures, thus quantifying the success rate by measures such as the $F_1$-score \cite{F1score} or overall accuracy. This strategy will be used to evaluate both the trained classifier performance with new/unknown time series data, as well as the accuracy during the learning phase.

For the evaluation of the performance during the learning phase, which is done after each classifier training session, the clusters that will emerge are checked versus the time series that were used as training material. Due to the unsupervised nature of the technique, and in order to match the output clusters with the original ground-truth labels, we select the highest-scoring $F_1$ score of all the permutations of the generated labels. A score $F_1=1.0$ corresponds to a perfect identification of the internal mapping that the TSC generated, and this mapping can then be used to validate the classifier with a new randomised time series.

\subsection{Optimised actuation sequence generation}
\label{rl_section}
As per \cref{eq:actuation_implementation}, there are infinitely many choices for the actuation vector $\tilde{u}\in[0,1]^N$. In reality, the actuators will not be able to respond to very low or very high percentages, and very similar actuation values would end up yielding the same results due to the actuators' finite response time. This is modelled with an inline low-pass filter in the actuation signals' path, between controller ouput and actuator input. In order to address this issue, as well as cut down on the size of the search space for choosing an actuation sequence, we consider the case where each actuation level $u_i \in (p_1, p_2, ..., p_M)$ with $p_i$ being evenly distributed predetermined percentages in the $[0, 1]$ range. This imposes that $M \geq 2$ so as to include at least complete turn-off and turn-on (i.e. $p_1=0$ and $p_M=1$)

Given the discrete nature of actuation values $u_i$, we can define an actuation sequence to be used with \cref{eq:state_space_representation_eq}, where each actuation vector will be applied for a constant duration. However, not all sequences will yield good results with respect to the classifier training/learning accuracy. In order for the classifier to learn how to identify the different inertia tensors, we need to provide responses that are adequately different in order to enable the $k$-means algorithm to be able to clearly identify the appropriate centroids. The performance of each sequence can only be evaluated by testing the classifier after training, thus a possible candidate sequence is first simulated with all the possible parameter combinations of the state space model, and then evaluated as as second step.

For sequences of small length and few actuation levels, all possible sequences can be tried, and subsequently compared against each other, in order to identify the optimal one. For each actuation application step with $M$ levels and $N$ thrusters, there are $M^N$ possible choices, and thus $L \times M^N$ overall actuation vectors where $L$ is the length of the actuation sequence. Given a minimum of 6 thrusters for complete controllability of the system, two levels of actuation (On/Off), even with $L=4$ discrete actuation steps, there are 256 possible actuation vectors, and the number increases as more actuation levels are introduced.

To solve this problem for longer sequences and more actuation levels, a reinforcement learning approach is applied using the PPO algorithm \cite{PPOpaper} implemented using the Stable Baselines\footnote{\url{https://github.com/DLR-RM/stable-baselines3}} library. The implemented model learns by trying different actuation vectors and is given a reward based on the performance of its selections. Thus, a policy is generated within the RL model such that for any given state it selects the next actions that maximise the reward, i.e. the actuation sequences that maximise the accuracy of identification. 

To completely decouple the state of the spacecraft from the actuation policy, the observation space for the RL model is defined as the history of actuation. Given a target actuation sequence of length $L$, this means that one training episode will consist of applying the sequence $\mathcal{O}=(\tilde{u}_1, \tilde{u}_2, ..., \tilde{u}_L)$, and the state or observation that is used within the RL agent is depicted as this sequence. Thus, after having trained the agent, we can stochastically determine what the next move should be after any step $\tilde{u}_i$ given the already used actuation up to that point so as to maximise the reward value.

For the simulation, a custom environment has been created to be used with the Gymnasium library\footnote{\url{https://gymnasium.farama.org/}} from OpenAI/Farama Foundation, which incorporates the following key elements:
\begin{itemize}
  \item Simultaneously evaluate multiple dynamics responses under the same actuation from different inertia tensors
  \item Use $\tilde{u}$ vectors as defined in \cref{eq:actuation_implementation} for the system actuation
  \item Time-stepping \cref{eq:state_space_representation_eq} with a RK4 solver for increased accuracy
  \item Update of RL agent policy by evaluation of the performance of the clustering algorithm during every step
\end{itemize}

The reward function that is used for the training of the RL agent during each training environment step $t_i$ for a sequence of length $L$ is as follows:

\[ 
\mathcal{R}_{step}= \left\{
\begin{array}{ll}
      -1 + 3 \alpha & \text{when } i<L \\
      - ||\bar{\omega}|| & \text{when } i=L \\
\end{array} 
\right. 
\]
where $\alpha\in[0, 1]$ is the classifier training evaluation score ($F_1$ score), and $\bar{\omega}$ is the angular velocity of the spacecraft. The last step's reward is used to promote sequences that produce the least overall disturbance to the spacecraft dynamics.

The hyperparameters used in PPO are as described in \cref{tab:hyperparameters}:

\begin{table}[h]\renewcommand{\arraystretch}{1.2}
\begin{center}
\begin{tabular}{c || c || c || c} 
\hline\hline
Parameter & Value & Parameter & Value\\
\hline\hline
n steps & $2048$ & learning rate & $0.0003$ \\
batch size & $64$ & gamma & $0.99$ \\
gae lambda & $0.95$ & clip range & $0.2$ \\
ent coef & $0.05$ & vf coef & $0.5$ \\
max grad norm & $0.5$ \\
\hline\hline
\end{tabular}
\caption{PPO Training hyperparameters}
\label{tab:hyperparameters}
\end{center}
\end{table}

\section{Results}

A payload deployer scenario is used to validate the proposed approach. A base inertia tensor has been selected as $I_0=diag(100, 60, 90) kg m^2$ and $4$ variations have been generated by $I_n=I_0 - {\delta}I$ with a uniform distribution $||{\delta}I||\in(0.05I_0, 0.15I_0)$, to model the state of different payloads having been deployed, which would alter the spacecraft's inertia tensor. The spacecraft is modelled as having 5 actuation levels evenly distributed between $0$ and $1.0$, and an actuation matrix A as follows, where $T=2 Nm$ is the nominal torque value
\begin{equation}
    A = T\begin{bmatrix}
        1 & 0 & 0 & -1 & 0 & 0 \\
        0 & 1 & 0 & 0 & -1 & 0 \\ 
        0 & 0 & 1 & 0 & 0 & -1 \\
    \end{bmatrix}
\end{equation}

\subsection{Actuation sequence optimisation}
\begin{figure}[h]
    \centering
    \includegraphics[width=.95\columnwidth]{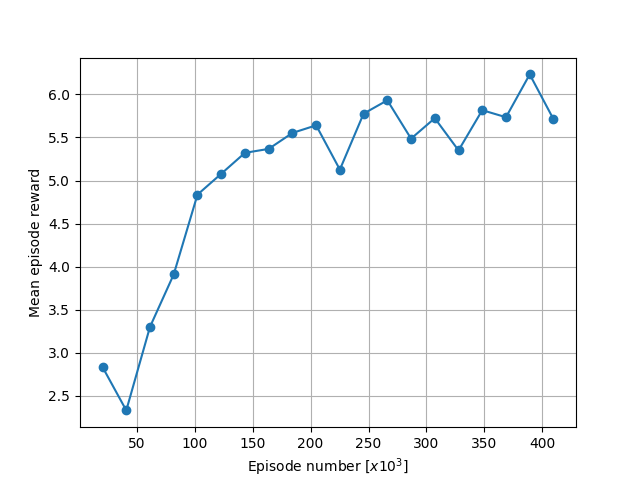}
	\caption{RL agent mean reward vs training episode}
	\label{fig:training_progress_reward}
\end{figure}

During the training session, the model's learning progress with respect to the pulse optimisation can be seen in \cref{fig:training_progress_reward}. The increasing value of the mean reward per episode demonstrates that the RL model is constantly optimising the policy up until the end of the learning phase, with the deviations from increasing reward values attributed to the exploratory nature of the algorithm in an attempt to navigate around possible local minima.

\subsection{Classifier robustness verification}
\begin{figure}[h]
    \centering
    \includegraphics[width=.95\columnwidth]{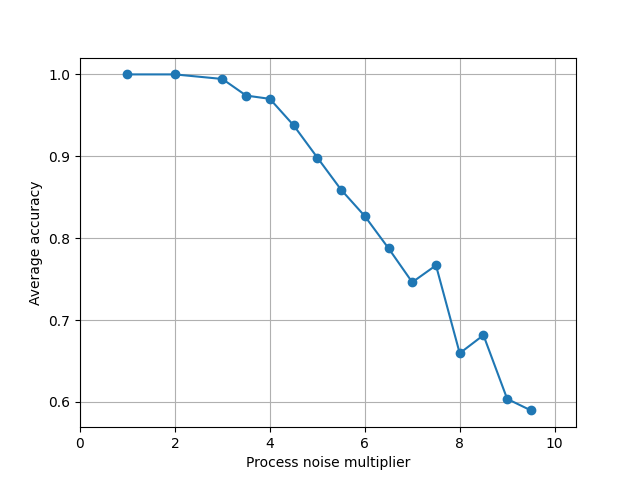}
	\caption{Classifier accuracy vs process noise}
	\label{fig:accuracy_vs_process_noise}
\end{figure}

\begin{figure}[h]
    \centering
    \includegraphics[width=.95\columnwidth]{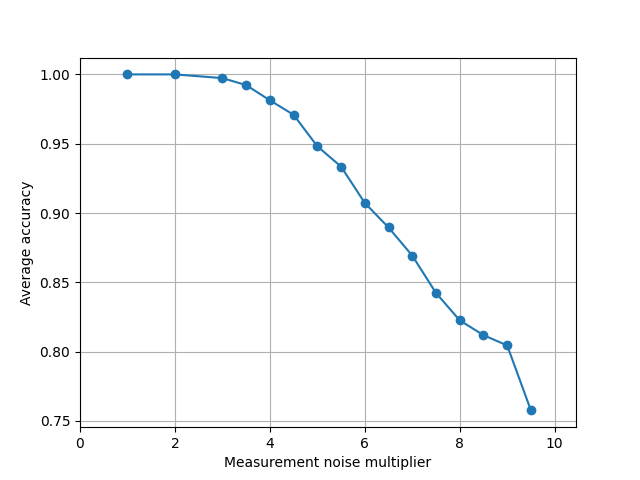}
	\caption{Classifier accuracy vs measurement noise}
	\label{fig:accuracy_vs_measurement_noise}
\end{figure}

In order to verify the classifier's accuracy, after the completion of the TSC model training, it may be tested against simulations of the same spacecraft models with similar noise and disturbance. In order to verify the robustness, the same classifier is tasked with identifying the spacecraft model when disturbance and noise levels increase by stepping the noise levels to multiples of the values for the training dataset. The results of this analysis can be seen in \cref{fig:accuracy_vs_process_noise} and \cref{fig:accuracy_vs_measurement_noise}.

These results show how in this scenario for commonly expected levels of noise in the dynamics and measurements, up to 3 times the values, we obtain perfect accuracy in the inertia tensor estimation. For very high levels of noise around 6 times the measured values, the accuracy levels decrease, but still remain above 80\% and 90\% for process and measurement noise.

\section{Conclusions}
In this work, a typical problem of state estimation was reshaped as a machine-learning solvable problem for a particular subset of spacecraft types and missions. The feasibility of the proposed approach was demonstrated by testing the time series clustering classifier with a realistic simulation model, which yielded promising results in the simulated scenario that encourage further research into the technique. 

Moreover, a degree of robustness was shown by testing the classifier in increasingly more difficult cases, both with sensor data of worse quality than the training dataset, and greater system disturbances. The classifier, partly due to the training regime with noisy data and partly due to the usage of lenient metrics with respect to time shift, exhibits some degree of robustness in both cases.

Finally, the choice of optimising the actuation profile with reinforcement learning via the PPO algorithm was proven successful, taking into consideration the training results which directly reflect the classifier performance increasing during training.

Given the promising early results, the proposed approach will be studied with more complicated systems that produce inherent disturbances. Possible candidates for this are complex structures with inherent flexibility, or more accurate single spacecraft models. In the case of complex structures, assuming distributed control and sensing capabilities, more data series would be available for the training phase and more actuation channels would be available for the RL optimisation of thruster firings. In the case of more accurate single spacecraft models, liquid sloshing effects on the dynamics (payload, or fuel) would provide for a good disturbance source to further validate the approach, and help identify any possible limitations.

\printbibliography
\addcontentsline{toc}{section}{References}

\end{document}